\documentclass[12pt]{article}
\usepackage{amssymb,amsbsy,latexsym}

\newcommand{\mbf}[1]{{\boldsymbol {#1} }}
\newcommand{\dd}{{\rm d}}
\newcommand{\ga}{\gamma}
\newcommand{\vzero}{\underline{{\mbf 0}}}
\newcommand{\ccr}[2]{\left[{#1},{#2}\right]}

\newcommand{\pp}{V}
\newcommand{\ee}[1]{\, {\rm e}^{#1}}
\newcommand{\ii}{{\rm i}}
\newcommand{\f}{\frac}
\newcommand{\Om}{\Omega}

\newcommand{\vx}{{\bf x}}
\newcommand{\vy}{{\bf y}}

\newcommand{\vn}{{\bf n}}

\newcommand{\vP}{{\bf P}}

\newcommand{\vE}{{\bf E}}
\newcommand{\vmu}{\underline{{\mbf \mu}}}

\newcommand{\Ref}[1]{(\ref{#1})}

\renewcommand{\b}{{\hat \rho}}
\newcommand{\U}{{\rm U}}
\newcommand{\sgn}{{\rm sgn}}
\newcommand{\eps}{\varepsilon}

\newcommand{\xxa}{\stackrel {\scriptscriptstyle \times}{\scriptscriptstyle \times} \!}
\newcommand{\xxe}{\! \stackrel {\scriptscriptstyle \times}{\scriptscriptstyle \times}}

\newcommand{\half}{\mbox{$\frac{1}{2}$}}

\newcommand{\Z}{{\mathbb Z}}

\newcommand{\cP}{{\cal P}}
\newcommand{\cH}{{\cal H}}
\newcommand{\cF}{{\cal F}}
\newcommand{\cE}{{\cal E}}

\newcommand{\eq}{\begin{equation}}
\newcommand{\eqend}{\end{equation}}
\newcommand{\eqa}{\begin{eqnarray}}
\newcommand{\nonueqa}{\begin{eqnarray*}}
\newcommand{\eqaend}{\end{eqnarray}}
\newcommand{\nonueqaend}{\end{eqnarray*}}
\newcommand{\nonu}{\nonumber \\ \nopagebreak}
\newcommand{\bma}[1]{\begin{array}{#1}}
\newcommand{\ema}{\end{array}}
\newcommand{\bc}{\begin{center}}
\newcommand{\ec}{\end{center}}

\begin{document}
\begin{flushright}
February 7, 2001
\end{flushright}
\vspace{.4cm}

\begin{center}

{\Large \bf Anyons and the elliptic
Calogero-Sutherland model}\\

\vspace{1 cm}

{\large Edwin Langmann}\\
\vspace{0.3 cm}
{\em Theoretical Physics, Royal Institute of Technology, S-10044 Sweden}\\
\end{center}

\begin{abstract} 

We obtain a second quantization of the elliptic Calogero-Sutherland
(eCS) model by constructing a quantum field theory model of anyons on
a circle and at a finite temperature.  This yields a remarkable
identity involving anyon correlation functions and providing an
algorithm for solving of the eCS model. The eigenfunctions obtained 
define an elliptic generalization of the Jack polynomials.

%\bigskip
%\noindent {MSC: 5Q58, 81T40, 17B69}

\end{abstract}

%%%%%%%%%%%%%%%%%%%%%%%%%%%%%%%%%%%%%%%%%%%%%%%%%%%%%%%%%%%%%%%%%%%%%%%%

There is an interesting class of integrable many body systems in one
dimensions which is known as Calogero-Moser-Sutherland systems
\cite{C,Su,M,OP}. These systems describe an arbitrary number of
identical particles interacting with a two-body potential which, in
the general case, is a Weierstrass elliptic function.  Apart from
their purely mathematical significance, these systems are also of
considerable physical interest as they are relevant to remarkably many
different topics including fractional statistics and anyons, quantum
chaos, QCD, and two dimensional quantum gravity (reviews of the first
two resp.\ three latter topics are, e.g., \cite{Poly} resp.\
\cite{Guhr}).

In the important limiting cases where the elliptic two-body potential
become rational and trigonometric, algorithms to solve the quantum
version of these systems were discovered by Calogero \cite{C} and
Sutherland \cite{Su}, and the mathematical properties of the solutions
thus obtained have been studied extensively (see e.g.\ \cite{F,Jack}).
In this letter we give a concise description of various novel results
on the elliptic Calogero-Sutherland (eCS) model, culminating in an
algorithm which provides a generalization of the above-mentioned
solution of Sutherland (detailed proofs with technical details will be
published elsewhere \cite{EL1,EL2}). In particular, we find a second
quantization of the eCS model as quantum field theory model of anyons
on a circle and at finite temperature, generalizing recent results on
the Sutherland model and zero temperature anyons \cite{sq,CL}.  We
then use this second quantization to prove a remarkable identity [Eq.\
\Ref{remarkable} below, which is a particular regularization of the
identity in Eqs.\ \Ref{rem1}--\Ref{rem2}] which finally yields an
algorithm to construct eigenfunctions and eigenvalues of the eCS
model. In the trigonometric limit our algorithm simplifies to one
which is equivalent to Sutherland's (i.e., it is different
but equally simple and yields the same solution) \cite{EL2}. To our
knowledge, solutions of the eigenvalue equation of the eCS Hamiltonian
were previously known only for integer values of the parameter
$\lambda$ defined in Eqs.\ \Ref{eCS} and \Ref{ga} below
\cite{WW,Lame,Felder}, whereas we do not impose such a restriction.

It is worth noting that in the two particle case, the eigenvalue
equation of the eCS is also known as {\em Lam\'e's equation} which was
studied extensively at the end of the 19$^{th}$ century (see
\cite{WW}) and more recently in \cite{Lame}. Results on the
$N$-particle generalization of this where obtained in \cite{Felder}
using quantum field theory techniques which, however, are different
from ours. Our approach was inspired by the relation of the Sutherland
model and the fractional quantum Hall effect (FQHE) \cite{FQHE}: it is
known that the edge excitation in FQHE states can be described by a 1D
quantum field theory model of composite fermions \cite{Wen} which is
identical with the anyon model used to construct the second
quantization of the Sutherland model \cite{sq,CL} at odd-integer
values of the parameter $\lambda$ where the anyons become (composite)
fermions [see Eq.\ \Ref{exc} below]. The second quantization mentioned
above \cite{sq,CL} therefore provides a direct link between the
Sutherland model and the FQHE. In this letter we find that the
generalization of the former to the elliptic case corresponds to a
natural generalization of the latter: going from zero- to finite
temperature.

We now fix our notation. The eCS Hamiltonian is defined by 
\eq
\label{eCS}
H_N = - \sum_{j=1}^N \frac{\partial^2}{\partial x_j^2} + \ga
\sum_{j<k} \pp(x_j-x_k) \eqend where $-\pi\leq x_j\leq \pi$ are
coordinates on a circle, $\ga>-1/2$ is the coupling constant,
$j,k=1,\ldots,N$, and the interaction potential is \eq
\label{pp}
\pp(r) \, = -\f{\partial^2}{\partial r^2} \log \vartheta_1 (\half r)
\eqend with the Jacobi theta function $\vartheta_1(r)= const.\
\sin(r)\prod_{n=1}^\infty [1-2q^{2n}\cos(2 r) + q^4]$ \cite{GR}.
Note that $\pp(r)$ is equal, up to an additive constant, to the
Weierstrass elliptic function $\wp(r)$ with periods $2\pi$ and $\ii
\beta$ where 
%$q=\ee{-\beta/2}$ 
$q = \exp(-\beta/2)$ \cite{GR}. The particle number $N$ is arbitrary
and $\beta>0$.  The Sutherland model corresponds to the limiting case
$q=0$ where $V(r)= (1/4) \sin^{-2}(r/2)$.  Of course, the differential
operator in Eq.\ \Ref{eCS} does not specify a unique quantum
mechanical Hamiltonian (i.e., it has many different self-adjoint
extensions), but our approach will automatically specify a unique
self-adjoint operator which, for $q=0$, coincides (essentially) with
the one diagonalized by Sutherland \cite{EL2}. (This self-adjoint
extension is particularly `nice' and, for $\ga >0$, essentially the
Friedrich's extension \cite{RS}.)
 
We now describe the construction of anyons using 1D chiral bosons
\cite{Wen} which can be made mathematically precise using the
representation theory of the loop group $Map(S^1;\U(1))$ \cite{CL}.
We recall that formally, an anyon field $\phi(x)$, parametrized by a
coordinate $-\pi\leq x\leq \pi$ on the circle, is characterized by the
exchange relations $\phi(x)\phi(y)=\ee{\pm
\ii\pi\lambda}\phi(y)\phi(x)$ where $\lambda>0$ is the so-called
statistics parameter. To construct such anyons we start with boson
operators $\b(n)$, $n$ integers, together with an invertible operator
$R$ and obeying the relations \eqa
\label{KM}
\ccr{\b(m)}{\b(n)}=m\delta_{m,-n}, \quad \ccr{\b(n)}{R} =
\delta_{n,0}R \: .  \eqaend It is natural to interpret $Q=\b(0)$ as
charge operator and $R$ as charge rising operator.  The standard
representation of this algebra is an irreducible highest weight
representation on a fermion Fock space (see e.g.\ \cite{CL}).  The
representation we use is different and will be specified further
below. In whatever representation, one can define operators \eq
K_\eps(x) = \sum_{n\neq 0} \f{1}{\ii n}\b(n)\ee{\ii nx}\ee{-|n|\eps}
\, , \eqend and then the anyons field can be defined as \eq
\phi_\eps(x) = \, \ee{-\ii\lambda Qx/2}\, R\, \ee{- \ii\lambda Qx/2}
\xxa \ee{ -\ii\sqrt{\lambda} K_\eps(x) } \xxe \eqend where $\eps>0$ is
a regularization parameter and $\xxa \cdot \xxe$ means normal ordering
which depends on the representation and will be specified below; for
now it is enough to know that $\xxa \cdot \xxe$ amounts to
multiplication with some $\eps$-dependent constant. We stress that
introducing this parameter $\eps$ is a convenient technical tool for
us which takes care of all the (ultraviolet) divergences which
otherwise would appear: for $\eps>0$ the quantum field $\phi_\eps(x)$
are well-defined operators which can be multiplied without difficulty.
Eventually we are interested in the limit $\eps\downarrow 0$ which is
singular (since the $\phi(x)=\phi_0(x)$ are operator valued
distributions), but we will be able to take this limit at a latter
point without difficulty.  Using Eq.\ \Ref{KM} and the Hausdorff
formula (i.e., $\ee{a}\ee{b}=\ee{-[a,b]}\ee{b}\ee{a}$ for operators
$a$ and $b$ such that $[a,b]$ commutes with $a$ and $b$), we obtain
exchange relations \eq
\label{exc}
\phi_\eps(x)\phi_{\eps'}(y) = \ee{-i\pi\lambda \sgn_{\eps+\eps'}(x-y)}
\phi_{\eps'}(y)\phi_\eps(x) \eqend where $ \sgn_\eps(x) = (1/\pi) [ x
+ \sum_{n\neq 0} \ee{in x -|n|\eps}/(\ii n) ] $ is a regularized sign
function on the circle. This shows that the $\phi_\eps(x)$ is indeed
a regularized anyon field operator with statistics parameter
$\lambda$.

We now specify the representation we are using and construct
our representation space $\cF$. For that we introduce 
two commuting copies of the algebra in Eq.\ \Ref{KM}, i.e.,  
$\ccr{\b_A(m)}{\b_B(n)} = m\delta_{m,-n}\delta_{A,B}$ for $A,B=1,2$, 
and similarly for $R_{1,2}$. Then $\cF$ is the Hilbert space generated
by these operators from a highest weight vector (vacuum) $\Om$ such that  
$\b_{A}(n)\Om=0$ and $\b_A(n)^*=\b_A(-n)$ for all $n\geq 0$ and 
that the $R_A$ are unitary operators such that
$<\Om, R_A^{m}\Om>=\delta_{m,0}$ for all integers $m$ ($*$ 
is the Hilbert space adjoint and $<\cdot,\cdot>$ 
the inner product). We then set
\eq
\b(n) = c_{|n|} \b_1(n) + s_{|n|} \b_2(-n) \quad
\forall n\neq 0 , 
\eqend
and $\b(0)=\b_1(0)$ and 
$R=R_1$, and it is easy to see that this gives a representation 
of the relations in Eq.\ \Ref{KM} provided that
$c_n^2-s_n^2=1$ for all $n=1,2,\ldots$. In particular we choose
\eq
\label{expl}
c_n = \left( \f{1}{1-q^{2n}}\right)^{1/2} ,\quad s_n = \left(
\f{q^{2n}}{1-q^{2n}}\right)^{1/2} \eqend 
with 
%$q=\ee{-\beta/2}$ 
$q=\exp{(-\beta/2)}$ 
and
$\beta>0$. One can prove that this is (essentially) the finite
temperature representation for the many particle Hamiltonian
$\cH_0=\sum_n\xxa \b(-n)\b(n)/2\xxe$ constructed by the usual trick of
doubling the degrees of freedom (see e.g.\ \cite{CHa}), and the
parameter $\beta$ corresponds to the inverse temperature
\cite{EL1}. To define the normal ordering prescription for anyon
operators we determine the creation- and annihilation parts of the
operator $K_\eps(x)$ such that $K=K^+ +K^-$, $K^- \Omega = 0$, and
$(K^-)^*=K^+$, i.e., $$ K_\eps^\pm(x) =\mp\sum_{n=1}^\infty  \f{1}{\ii n}[
c_n\b_1(\mp n)\ee{\mp \ii nx} - s_n \b_2(\mp n)\ee{\pm \ii
nx}]\ee{-n\eps} \, . $$ 
We will need the commutators of the latter operators. 
A straightforward computation yields
\eq
\label{KpKm}
\ccr{K^-_\eps(x)}{K^+_{\eps'}(y)} = C_{\eps+\eps'}(x-y) \eqend 
where $C_\eps(r)=\sum_{n=1}^\infty ( c_n^2
\ee{\ii nr } + s_n^2 \ee{-\ii n r})\ee{-n\eps} /n$.  Inserting Eq.\
\Ref{expl} and expanding $1/(1-q^{2n})$ in geometric series we obtain
$$C_\eps(r) = -\log[ b_{\eps}(r) \exp{(\ii r/2)}]$$ with \eq
\label{b}
b_\eps(r) =
 -2\ii \ee{-\eps/2}
\sin(\f{r+i\eps}{2}) 
%\nonu \times 
\prod_{n=1}^\infty [1-2q^{2n}\ee{-\eps}\cos(r)+q^{4n}\ee{-2\eps} ] \:
.  \eqend Note that $b_\eps(r)$ is essentially a regularization of the
above-mentioned Jacoby theta function, $b_0(r)= const.\,
\vartheta_1(r/2)$.  We now can define normal ordering,
$$\xxa \ee{ -\ii\sqrt{\lambda} K_\eps(x) } \xxe \, =
\ee{-\ii\sqrt{\lambda} K^+_\eps(x) } \ee{-\ii\sqrt{\lambda}
K^-_\eps(x) } , $$ and a simple computation shows that this amounts to
a multiplication with the constant $b_{2\eps}(0)^{-\lambda/2}$.  This
completes our construction of the anyon model.  One can now compute
all anyon correlation functions by using the Hausdorff formula and
Eq.\ \Ref{KpKm}, for example the function \eqa
\label{FN}
F^{\eps,\eps'}(\vx;\vy)\equiv 
F^{\eps,\eps'}(x_1,\ldots, x_N ; y_1,\ldots, y_N) :\, 
= \nonu =  
\left<\Omega, \phi_{\eps}(x_N)^*
\cdots \phi_{\eps}(x_1)^* \phi_{\eps'}(y_1)\cdots 
\phi_{\eps'}(y_N)
 \Omega \right> = \nonu
= \f{ \prod_{1\leq j<k\leq N} b_{2\eps}(x_{k}-x_j)^{\lambda} 
b_{2\eps'}(y_{j}-y_{k})^{\lambda}}{\prod_{j,k=1}^N
b_{\eps+\eps'}(x_j-y_k)^{\lambda}}  
\eqaend
which will play an important role further below. 

Having set the stage, we now can
describe the second quantization of the eCS Hamiltonians 
$H_N$ in Eq.\ \Ref{eCS} 
and how this leads to a remarkable 
identity which will be the starting point for our solution 
algorithm: we found a self-adjoint operator 
$\cH$ on $\cF$ such that the commutator of $\cH$ with a product
of $N$ anyon operators, 
\eq
\label{PhiN}
\Phi^{N}_\eps (\vx) = \phi^{\nu}_{\eps}(x_1)\cdots
\phi^{\nu}_{\eps}(x_N)\, , \eqend is essentially equal to $H_N
\Phi^{N}_\eps (\vx)$ where $H_N$ is the eCS Hamiltonian with a
coupling constant determined by the statistics parameter as follows,
\eq
\label{ga}
\ga = 2\lambda(\lambda-1) \, . 
\eqend 
To be more precise: this operator $\cH$ obeys the relations
\eq
\label{cc}
[\cH,\Phi^{N}_\eps (\vx)]\Om \simeq H_N^\eps \Phi^{N}_\eps (\vx)\Om
\eqend with $H^{2\eps}_N$ as in Eq.\ \Ref{eCS} but with $V(r)$
replaced by the regularized potential $V_{2\eps}(r)=-\partial^2 \log
b_{2\eps}(r)/\partial r^2$, and `$\simeq$' means `equal in the limit
$\eps\downarrow 0$'. We note that it is surprisingly simple to
construct this operator $\cH$ by following the arguments for the
Sutherland model \cite{CL}. It has the form \eqa \cH =
\frac{\sqrt{\lambda}}{3} \sum_{m,n} \xxa \b(m+n)\b(-m)\b(-n) \xxe + \nonu + 
(1-\lambda)
\sum_{n>0} n[ \b_1(-n)\b_1(n) + \b_2(-n)\b_2(n) ] +\ldots \eqaend where
the dots indicate less important terms proportional to $Q\cH_0$, $Q^3$
and $Q$ \cite{EL1}. It is interesting to note that the relations in
Eq.\ \Ref{cc} can be established without using Eq.\ \Ref{expl}, i.e.,
the second quantization $\cH$ obeying Eq.\ \Ref{cc} exists for a much
larger class of Hamiltonians given in Eq.\ \Ref{eCS} with
$V(r)=V_0(r)$, \eq
\label{Veps}
V_\eps(r)= -\f{\partial^2}{\partial r^2}  
\sum_{n=1}^\infty \f{1}{n}
\left(  \ee{\ii n r}
+ s_n^2 [ \ee{\ii n r} + \ee{-\ii n r}] \right) \ee{-n\eps} \: . 
\eqend
However, as we will see, to obtain an solution algorithm one also needs 
\eq
\label{id}
<\Omega, [\cH, \Phi^{N}_{\eps}(\vx)^* \Phi^{N}_{\eps'}(\vy)] \Omega>\, 
= 0
\eqend
(note that this identity is trivial in the Sutherland case
where  $\cH\Om=0$, but this no longer holds in general): 
we proved that this identity holds true
if and only if 
\eq
c_m^2c_n^2s_{m+n}^2 = s_m^2 s_n^2 c_{m+n}^2 
\quad \forall m,n=1,2,\ldots  
\eqend
which restricts us to $c_n$ and $s_n$ as in 
Eq.\ \Ref{expl} and thus interaction potentials
which are Weierstrass elliptic functions (this proof 
is by a straightforward computation using the 
explicit formula for $\cH$ given above \cite{EL1}).  
We now compute the vacuum expectation value of
the (trivial) identity   
$$[\cH, \Phi^{N}_{\eps}(\vx)^*\Phi^{N}_{\eps'}(\vy) ] = 
 - [\cH,\Phi^{N}_{\eps}(\vx)]^* \Phi^{N}_{\eps'}(\vy) 
+  \Phi^{N}_{\eps}(\vx)^*[\cH,\Phi^{N}_{\eps'}(\vy)]$$
using Eq.\ \Ref{id}. Inserting Eq.\ \Ref{cc} twice we obtain 
\eq
\label{remarkable}
\overline{H_N^{2\eps}(\vx)} F^{\eps,\eps'}(\vx;\vy) 
\simeq
H_N^{2\eps'}(\vy)F^{\eps,\eps'}(\vx;\vy)
\eqend 
with $F^{\eps,\eps'}(\vx;\vy) \, = \, {<\Omega,}
\Phi^{N}_{\eps}(\vx)^*\Phi^{N}_{\eps'}(\vy)\Omega> $ the anyon
correlation function defined and computed in Eq.\ \Ref{FN} above and
regularized eCS Hamiltonians acting on different variables as
indicated (the bar means complex conjugation). This is our remarkable
identity and a main result of this letter. 

We now show how Eqs.\ \Ref{remarkable} and \Ref{FN} can be used
to construct eigenfunctions and eigenvalues of the eCS Hamiltonian. 
The idea is to take the Fourier transform
of Eq.\ \Ref{remarkable}, i.e., apply to it $(2\pi)^{-N}\int
\dd^N\vy\,\ee{\ii \vP\cdot \vy}$ (the integration is over
$-\pi\leq y_j\leq \pi$, of course), and then  
take the limits $\eps,\eps'\downarrow 0$. 
To determine the possible values for 
the Fourier modes $P_j$ we observe that $b_\eps(r)^\lambda$ 
changes  by a factor $\ee{\mp \ii \pi\lambda}$ under $r\to r \pm 2\pi$.  
Thus the function $F^{\eps,\eps'}(\vx;\vy)$ 
changes by a factor $\ee{-\ii \pi (2N-2j+1)\lambda }$ under 
$y_j\to y_j+2\pi$. 
The $P_j$ need to be such that $\ee{\ii \vP \cdot \vy} 
F^{\eps,\eps'}(\vx;\vy)$ is periodic, which enforces
$
P_j = n_j + (N-j+\half) \lambda$ with integers $n_j$. These 
are (essentially) the `quasi-momenta' known from the Sutherland 
model (up to a trivial shift due to a 
center-of-mass motion contained in our 
eigenfunctions; see Eq.\ \Ref{series} below\footnote{The interested 
reader can check that removing this center-of-mass 
motion amounts to changing the quasi-momenta as follows, 
$P_j\to n_j + \lambda(N+1-2j)/2$, 
and by relabeling $n_j\to n_{N+1-j}$ one obtains 
Sutherland's expression \cite{Su}.}). 
With that we obtain
\eqa
\label{abc}
H_{N} \hat F(\vx;\vn) = \cE_0(\vn) \hat F(\vx; \vn) -
\ga\sum_{j<k}\sum_{n=1}^\infty n \nonu \times \left[ c_n^2 \hat F(\vx; \vn
+ n\vE_{jk}) + s_n^2 \hat F(\vx; \vn- n\vE_{jk}) \right]
\eqaend with $(\vE_{jk})_\ell =
\delta_{j\ell}- \delta_{k\ell}$ for $\ell=1,\ldots,N$, and \eq
\cE_0(\vn) = \sum_{j=1}^N P_j^2 = \sum_{j=1}^N \left[ n_j +
(N-j+\half) \lambda \right]^2 \: , \eqend where the first term on the
r.h.s.\ in Eq.\ \Ref{abc} comes from the derivative term in the eCS
Hamiltonian and partial integration, and the second term comes from
the interaction terms which we evaluated using Eq.\ \Ref{Veps}.  The
function $\hat F(\vx;\vn)$ is the $\eps,\eps'\downarrow 0$-limit of
the Fourier transform of $F^{\eps,\eps'}(\vx;\vy)$, i.e., \eq
\label{a}
\hat F(\vx;\vn) = 
\cP(\vx;\vn) \, \Delta(\vx)\, \ee{ \sum_j \ii\lambda x_j}
\eqend
where 
\eq
\label{Delta}
\Delta(\vx) = 
\prod_{j<k} b_0(x_k-x_j)^\lambda
\eqend
and the symmetric, periodic function
\eq
\label{P}
\cP(\vx;\vn) = \lim_{\eps,\eps'\downarrow 0} \int \!
\f{\dd^N\vy}{(2\pi)^{N}} \,\ee{\ii \vn\cdot \vy} \f{ \prod_{j<k }
\check b_{2\eps}(y_j-y_k)^\lambda}{ \prod_{j,k} \check
b_{\eps+\eps'}(x_j-y_k)^\lambda } \eqend 
with
$ \check b_\eps(r) =
%\ee{ \ii r/2} b_\eps(r) = 
\left(1 - \ee{i r-\eps} \right)
\prod_{n=1}^\infty [1-2q^{2n}\ee{-\eps}\cos(r)+q^{4n}\ee{-2\eps} ]$;
the last factor in Eq.\ \Ref{a} describes an uninteresting
center-of-mass motion.
We defined the functions $\cP$ as a
particular regularization of singular integrals, but these integrals
can be computed explicitly by expanding in plane waves \cite{CL}.
Expanding also in powers on $q^2$ one obtains $ \cP =
\sum_{\ell=0}^\infty \cP^\ell \, q^{2\ell} $ where the $\cP^\ell$
are symmetric polynomials, i.e., their expansion in plane waves has
only a {\em finite} number of non-zero terms (this number of terms
diverges as $\ell$ goes to infinity, however).

Writing $\vmu=\sum_{j<k }\mu_{jk}\vE_{jk}$ 
with integer $\mu_{jk}$ and identifying the set of all
such $\vmu$ with $\Z^{N(N-1)/2}$ we now make the 
following ansatz for an eigenfunction, 
\eq
\label{ansatzN}
\psi(\vx) = \sum_{\vmu} 
\alpha(\vmu) \, \hat F(\vx;\vn+\vmu)
\eqend
(we suppress the common argument $\vn$ of $\psi$, $\alpha$, 
$\cE$ in the following). Then the equation 
\eq
\label{eigen}
H_N\psi = \cE \psi
\eqend 
implies  
\eq
\label{main0}
[\cE_0(\vn+\vmu) -\cE ]\, \alpha(\vmu) = 
\ga\sum_{j<k} 
\sum_{n=1}^\infty n [ c_n^2 \alpha(\vmu -n\vE_{jk}) + 
s_n^2 \alpha(\vmu +n\vE_{jk})]  \, . 
\eqend
To solve these equations we make the ansatz
\eq
\alpha(\vmu) = \sum_{\ell=0}^\infty \alpha_\ell(\vmu) \, q^{2\ell}\: , \quad 
\cE = \sum_{\ell=0}^\infty \cE_\ell \, q^{2\ell} \: . 
\eqend
Using $s_n^2=\sum_{m=1}^\infty q^{2nm}$ and $c_n^2=1+s_n^2$ we get
\eqa
\label{main}
[\cE_0(\vn+\vmu) -\cE_0]\, \alpha_\ell(\vmu) -
\sum_{m=1}^\ell \cE_m \alpha_{\ell-m}(\vmu)
= \nonu = 
\ga\sum_{j<k} 
\sum_{n=1}^\infty n
\alpha_\ell(\vmu -n\vE_{jk}) + 
 \ga \sum_{j<k} \sum_{\stackrel{n,m>0}{nm\leq \ell}} \nonu \times  
n[
\alpha_{\ell-nm}(\vmu-n\vE_{jk}) + \alpha_{\ell-nm}(\vmu+n\vE_{jk}) 
] \: . 
\eqaend
Eqs.\ \Ref{a}--\Ref{main} constitute our algorithm to
solve the eCS model. We can restrict ourselves to 
$(\ell,\vmu)$ such that 
\eq
\label{restr}
\mu_{jk} \geq -\ell \quad \forall j<k \eqend [i.e.\ for other
$(\ell,\vmu)$ we set $\alpha_\ell(\vmu)=0$] and determine the
$\alpha_\ell(\vmu)$ and $\cE_\ell$ from Eq.\ \Ref{main} recursively.
It is important to note that this recursive procedure has {\em
triangular structure}, i.e., amounts to diagonalizing triangular
matrices. To see that we observe that there is a natural partial
ordering, $(\ell',\vmu')<(\ell,\vmu)$ if
$$
\ell'<\ell \mbox{ or ($\ell'=\ell$ and $\mu'_{jk}\leq \mu_{jk}\quad 
\forall j<k$ and $\vmu'\neq\vmu$), } 
$$
and for fixed $(\ell,\vmu)$ there is only finitely many
$(\ell',\vmu')$ with $(\ell',\vmu')<(\ell,\vmu)$. For $\vmu=\vzero$
and $\ell=0$ we get $\cE_0=\cE_0(\vn)$, and $\alpha_0(\vzero)$ remains
a free parameter.  For $\ell>0$ and $\vmu=\vzero$ we get an equation
determining $\cE_\ell$ as a sum of finitely many terms depending only
on the $\alpha_{\ell'}(\vmu')$ and $\cE_{\ell'}$ with
$(\ell',\vmu')<(\ell,\vzero)$, and $\alpha_\ell(\vzero)$ remains
undetermined.  For non-zero $\vmu$, there are two different cases. If
there is a resonance, i.e., if the factor \eqa \cE_0(\vn +
\vmu)-\cE_0(\vn ) = 2 \sum_{j<k} \left[n_j-n_k + (k-j)\lambda\right]
\mu_{jk} + \nonu + \sum_j \left( \sum_{k<j}\mu_{k j} - \sum_{k>j}
\mu_{jk} \right) ^2 \eqaend vanishes, we get a linear equation
constraining the previously undetermined $\alpha_{\ell'}(\vmu')$ for
$(\ell',\vmu')<(\ell,\vmu)$, and $\alpha_\ell(\vmu)$ remains
undetermined.  In the generic case, i.e.\ if there is no resonance,
$\alpha_\ell(\vmu)$ is determined as a finite sum of terms depending
only on $\alpha_{\ell'}(\vmu')$ and $\cE_{\ell'}$ for $(\ell',\vmu') <
(\ell,\vmu)$. We thus can determine all $\alpha_\ell(\vmu)$
recursively, and at each order $\ell$ there remains only one of them
undetermined [generically this will be $\alpha_\ell(\vzero)$].  We
note that these free parameters correspond to the freedom of choosing
different normalizations of the eigenfunctions, i.e., changing them
amounts to multiplying the eigenfunction $\psi(\vx)$ by a
normalization constant which is a power series in $q^2$. Moreover, it
is known that in the Sutherland case one only needs to consider
$N$-tuples $\vn$ such that \eq
\label{hw}
n_1\geq n_2\geq \cdots \geq n_N\geq 0 \eqend 
since these already
provide a complete set of eigenfunctions \cite{CL,Jack}, and we expect
this is true also for $q\neq 0$.

It is clear that resonances make our algorithm somewhat more involved,
and it is therefore interesting to mention some cases where resonances
can be ruled out.  For example, there is never a resonance for $\vmu >
\vzero$ and $\vn$ as in Eq.\ \Ref{hw} \cite{CL}, and therefore
resonances do not occur in the Sutherland case $q=0$.  Moreover, for
$N=2$, it is easy to see that resonances can only occur if $\lambda$
is integer. However, for $N>2$, there are infinitely many resonances
which are independent of $\lambda$, e.g.\ for $N=3$ and $\vn$ such
that $n_1-2n_2+n_3=3\nu$ with integer $\nu$, one has a resonances for
all $\vmu$ such that $\mu_{13}=-\nu-\mu_{12}$ and $\mu_{23}=
2\nu+\mu_{12}$ ($\mu_{12}$ arbitrary integer), and for rational values
of $\lambda$, additional `coincidental' resonances (i.e., they depend
on $\lambda$) are to be expected. Thus, for $N=3$, resonances can be
ruled out for irrational $\lambda$ and $\vn$ such that
$(n_1-2n_2+n_3)/3$ is non-integer. Obviously, a more general analysis
of the occurrence and implications of resonances would be welcome.

The eigenfunctions which we get are of the form \eq
\label{series}
\psi(\vx)= J_{\vn} (\vx|q) \, \Delta(\vx)\,\ee{\sum_j \ii N \lambda
x_j} \eqend where we defined  $J_{\vn} (\vx|q) \equiv \sum_{\ell=0}^\infty
J_{\vn}^\ell (\vx)\, q^{2\ell}$ with 
$$J_{\vn}^\ell (\vx)= \sum_{\ell'=0}^\ell \sum_{\vmu}
\alpha_{\ell-\ell'}(\vmu)\, \cP^{\ell'}(\vx;\vn+\vmu)\, . $$ It is
interesting to note that the latter sums are always finite, i.e., one
can prove highest weight relations for the functions $\cP^\ell$ which
imply that there are only finitely many $\vmu$ obeying Eq.\
\Ref{restr} and such that $\cP^{\ell'}(\vx;\vn+\vmu)$, $0\leq
\ell'\leq \ell$, are different from zero \cite{EL2}.  Moreover, as already
mentioned, the $J_{\vn}(\vx|q)$ are uniquely determined up to
normalization. In the case $q=0$ our algorithm reduces to the one in
Ref.\ \cite{CL}, and the results there imply that the
$J_{\vn} (\vx|q=0) = J^0_{\vn}(\vx)$, $\vn$ obeying the condition in
Eq.\ \Ref{hw}, are proportional to the Jack polynomials \cite{Jack}.
It is therefore natural to regard the $J_{\vn} (\vx|q)$ as elliptic
generalization of the Jack polynomials.

Recently it was shown that the formal power series expansion of the
eigenfunctions of the eCS model in $q^2$ converge in the $L^2$-norm
sense \cite{Ko}. This result suggests that the formal power series
which we obtained actually converge, and in particular, that our
elliptic generalizations of the Jack polynomials are well-defined
symmetric functions. For $q=0$ it is known that they define a complete
orthogonal set of eigenfunctions \cite{Jack}, and we conjecture the
same is true also for finite $q$. Obviously, a more detailed
investigation of these functions would be welcome.

We conclude with a two remarks. Firstly, we note that the only result
from our anyons construction which we actually needed to get our
solution algorithm is the remarkable identity in Eq.\
\Ref{remarkable}. For $\eps=\eps'=0$ the latter reduces to the
following identity of elliptic functions, \eq
\label{rem1}
% H_N(\vx) F(\vy;\vx) = H_N(\vy) F(\vy;\vx) 
\sum_{j=1}^N 
\biggl( \f{\partial^2}{\partial x_j^2} -  
\f{\partial^2}{\partial y_j^2}
\biggr) F(\vx;\vy) = 
\sum_{j < k} 2\lambda(\lambda-1) \biggr[  \wp(x_j-x_k) 
- \wp(y_j-y_k) \biggr] F(\vx;\vy) 
\eqend where
\eq
\label{rem2}
F(\vx;\vy) = \f{ \prod_{1\leq j<k\leq N}
\vartheta_1(\half[x_{k}-x_j])^{\lambda}
\vartheta_1(\half[y_{j}-y_{k}])^{\lambda}}{\prod_{j,k=1}^N
\vartheta_1(\half[x_j-y_k])^{\lambda}}\, .  \eqend We suggest that {\em
it is this remarkable identity which distinguishes the quantum
Calogero-Moser-Sutherland models} [rather than the existence of a
product ground state function of the form as in Eq.\ \Ref{Delta}]
since it can be used to obtain an algorithm to solve the model and
holds for the elliptic case as well. After this work was completed we
found an elementary, direct proof of this based on the following
identity of Weierstrass elliptic functions,
$$
[\zeta(x) +\zeta(y) + \zeta(z)]^2 = \wp(x) + \wp(y) + \wp(z) \quad
\mbox{ if $x+y+z=0$} 
$$ (whose proof is given as an exercise on p.\ 446 in Ref.\
\cite{WW}) \cite{EL2}. Still, the generalization which we got from the
anyons contains important additional information: the
$\eps$-regularization is needed to fix a unique self-adjoint
extension of the differential operator $H_N$ in Eq.\
\Ref{eCS}. Finally, we note that it would be interesting to find an
argument to obtain from the identity in Eq.\ \Ref{remarkable} an
algorithm providing {\em finite} series representations of the
eigenfunctions, e.g., by finding a way to solve Eq.\ \Ref{main0}
non-perturbatively (i.e., without expanding in $q^2$).

%To summarize, we constructed a quantum field theory model of anyons on
%the circle and at finite temperature using a
%vertex operator construction. From that we obtained a second
%quantization of the elliptic Calogero-Sutherland system.  This allowed
%us to prove a remarkable identity involving elliptic functions and
%providing an algorithm for constructing the eigenfunctions and
%eigenvalues of the eCS Hamiltonian, for arbitrary couplings and
%particle numbers and in one-to-one correspondence with the complete
%solutions of the Sutherland model \cite{Su} which we recover in the
%trigonometric limit.
\bigskip

\begin{center}
{\bf Acknowledgements:}
\end{center}

I thank A.\ Carey and A.\ Polychronakos for their interest and helpful
discussions and K.~Takemura for pointing out to me Ref.\ \cite{Ko}. A
helpful remark of K.\ Fredenhagen is acknowledged. This work was
supported by the Swedish Natural Science Research Council (NFR).

%\end{article}

\end{document}